\documentclass[12pt,preprint]{emulateapj}

\begin{document}

\shorttitle{Comptonized X-rays from Millisecond Pulsars}
\shortauthors{Bogdanov, Grindlay, \& Rybicki}

\title{X-rays from Radio Millisecond Pulsars: Comptonized Thermal Radiation}

\author{Slavko Bogdanov, Jonathan E. Grindlay, and George B. Rybicki}
\affil{sbogdanov@cfa.harvard.edu, josh@cfa.harvard.edu, grybicki@cfa.harvard.edu \\
Harvard-Smithsonian Center for Astrophysics, 60 Garden Street, Cambridge, MA 02138}

\begin{abstract}

X-ray emission from many rotation-powered millisecond pulsars
(MSPs) is observed to be of predominantly thermal nature. In PSR
J0437--4715, the nearest MSP known, an additional faint power-law tail
is observed above 2.5 keV, commonly attributed to non-thermal
magnetospheric radiation. We propose that the hard emission in this
and other similar MSPs is instead due to weak Comptonization of the
thermal (blackbody or hydrogen atmosphere) polar cap emission by
energetic electrons/positrons of small optical depth in the pulsar
magnetosphere. This spectral model implies that all soft X-rays are of
purely thermal origin, which has profound implications in the study of
neutron star structure and fundamental pulsar physics.

\end{abstract}


\section{Introduction}

As the nearest and brightest rotation-powered millisecond pulsar
(MSP), PSR J0437-4715 \citep{John93,van01} has been the best studied
in the X-ray domain.  \textit{ROSAT}, \textit{Chandra}, and
\textit{XMM-Newton} observations \citep{Zavlin98,Zavlin02,Zavlin06}
have revealed that the emission in the 0.1-10 keV band consists of two
thermal components, and a faint power-law (PL) tail, with photon index
$\Gamma\sim 2$. The latter can only be discerned in the spectrum
beyond 2.5 keV.  Due to the limited spectral information above $\sim$3
keV, the nature of this hard X-ray component is unclear.

In this \textit{Letter}, we examine the plausibility of various
emission mechanisms capable of producing the hard X-rays observed from
J0437-4715 (henceforth J0437). The paper is organized as follows. In
\S2 we describe the observations used and their analysis while \S3
outlines various interpretations of the emission from J0437. In \S4 we
present the results of our analysis and offer a discussion and
conclusions in \S5.

\section{OBSERVATIONS AND DATA REDUCTION}

We have retrieved archival \textit{Chandra} and \textit{XMM-Newton}
observations of J0437 to investigate the X-ray emission from this MSP.
The \textit{Chandra} ACIS-S observations were performed on 2000 May
29-30 \citep[25.7 ks effective exposure, 3.9' off axis; see][for
details]{Zavlin02}. We first reprocessed the level 1 event files using
the most recent calibration release and the CIAO 3.3 software
package\footnote{Chandra Interactive Analysis of Observations,
available at http://asc.harvard.edu/ciao/}. The data were then
filtered to the nominal \textit{Chandra} energy band (0.3--8 keV). The
source counts were obtained from a 5'' circular region centered on the
radio position of the MSP. This circle encloses $>$90\% of the total
energy.

The \textit{XMM-Newton} data consist of a 70-ks observation with the
EPIC-MOS1/2 detectors in full-frame mode and EPIC-pn instrument in
fast timing mode obtained on 2002 October 9 \citep[see][]{Zavlin06}.
The raw data were reprocessed using the {\tt emchain} and {\tt
epchain} pipelines of the SAS 6.5.0 software\footnote{Science Analysis
Software, available at http://xmm.vilspa.esa.es/} and were screened
for instances of high background flares, which were subsequently
removed. In the case of MOS1/2, the source counts were extracted from
70'' circles, which enclose $>$90\% of the total energy. To minimize
background contamination the MOS1/2 data were restricted to the 0.3--8
keV band.  For the pn detector, the MSP counts were extracted from CCD
columns 28-39 and were filtered to 0.5--10 keV, where the detector
response is best known.

For all observations, the background was obtained from a source-free
region near the position of the MSP. The spectral fits were performed
in XSPEC 11.3.1\footnote{http://heasarc.gsfc.nasa.gov/docs/xanadu/xspec/}.

\section{ORIGIN OF THE HARD X-RAY TAIL}

The inherent faintness of the hard PL component from J0437 and the
limited bandwidth over which it can be observed pose an obstacle in
deciphering the origin of this emission.  Nonetheless, using the
available data we can by gain more insight by examining the
plausibility of the various mechanisms for production of the observed
hard X-ray tail.

\subsection{Non-thermal Emission}

The most widely accepted model for production of a PL spectrum in
pulsars involves non-thermal radiation by relativistic electrons and
positrons (e$^{\pm}$) in the pulsar magnetosphere.  The analyses by
\citet{Zavlin02} and \citet{Zavlin06} of \textit{Chandra} and
\textit{XMM-Newton} observations found that the hard X-ray tail seen
in J0437 is well described by a PL with spectral index
$\Gamma=2.0\pm0.4$.  Under the assumption that this emission is of
non-thermal magnetospheric origin, the PL is expected to extend from
the optical/UV and beyond the hard X-ray range ($>$10 keV).  However,
\textit{Hubble Space Telescope} FUV observations have revealed that
extrapolation of the powerlaw to lower energies is inconsistend with
the FUV flux from the MSP, with the lower bound $\Gamma\approx1.6$
only marginally consistent with the FUV flux \citep{Kar04}. The
measured FUV flux is otherwise consistent with that of $\sim$10$^5$ K
thermal emission from the whole NS surface.  The apparent discrepancy
casts doubt on the validity of the PL model.

\citet{Beck99} have proposed that the X-ray emission from J0437 and
other nearby MSPs (e.g.~PSRs J0030+0451 and J2124--3358) could be due
to non-thermal processes in the magnetosphere.  However, based on more
recent observations of these MSPs, we conclude that this interpretation
is untenable. If the emission were indeed of purely non-thermal
character, then such a broken PL spectrum grossly overestimates the
observed optical/UV fluxes \citep{Kop03,Mig03,Kar04} unless yet
another break exist below 0.1 keV.  No such behavior is observed in
the X-ray spectra of other pulsars, in particular the luminous MSPs
B1937+12 and B1821-24, which show purely non-thermal pulsed radiation
\citep[see e.g.,][]{Beck03,Min04}.

An alternative source of non-thermal X-rays is nebular synchrotron
emission from an intrabinary shock formed due to interaction between
the wind from the MSP and the companion star. Such a shock is believed
to be present in other binary MSPs, namely PSRs B1957+20
\citep{Stap03}, J0024--7204W \citep{Bog05}, and J1740--5340 (Grindlay
et al. in preparation). However, for J0437, the strength of the pulsar
wind $\dot{E}\sim4\times10^{33}$ ergs s$^{-1}$ and the orbital
separation\footnote{In \citet{Zavlin02} the semi-major axis of the MSP
($a_1$) is erroneously quoted as being the distance between the MSP
and its companion ($a=a_1+a_2$).} of $a=1.2\times10^{12}$ cm do not
favor this interpretation as the expected X-ray luminosity from the
resulting shock is substantially lower than observed.  Specifically,
assuming a radius $\sim$$1.5\times10^9$ cm for the 0.24 M$_{\odot}$
white dwarf (WD) companion \citep{van01}, we find that the total
incident power on the face of the WD from the pulsar wind is
$\sim$$10^{27}$ ergs s$^{-1}$, more than two orders of magnitude lower
than the observed luminosity of the hard emission ($L_X\approx 7
\times10^{29}$ ergs s$^{-1}$ for 0.1--7 keV). The large discrepancy
can be reduced in a more contrived scenario, involving a highly
anisotropic wind, confined to the orbital plane in a beam covering
$<$0.15\% of the sky. Alternatively, the companion may be bloated or
surrounded by a large tenuous envelope of gas that interacts with the
MSP wind. However, this assertion is in conflict with the observed
optical properties of J0437, which are entirely consistent with that
of a $\sim$0.2 M$_{\odot}$ He-WD \citep[see
e.g.,][]{Bay93}.  Therefore, we conclude that an intrabinary shock is
not a significant source of X-ray emission in the J0437 system.


\begin{figure}[!t]
\begin{center}
\includegraphics[angle=270, width=0.47\textwidth]{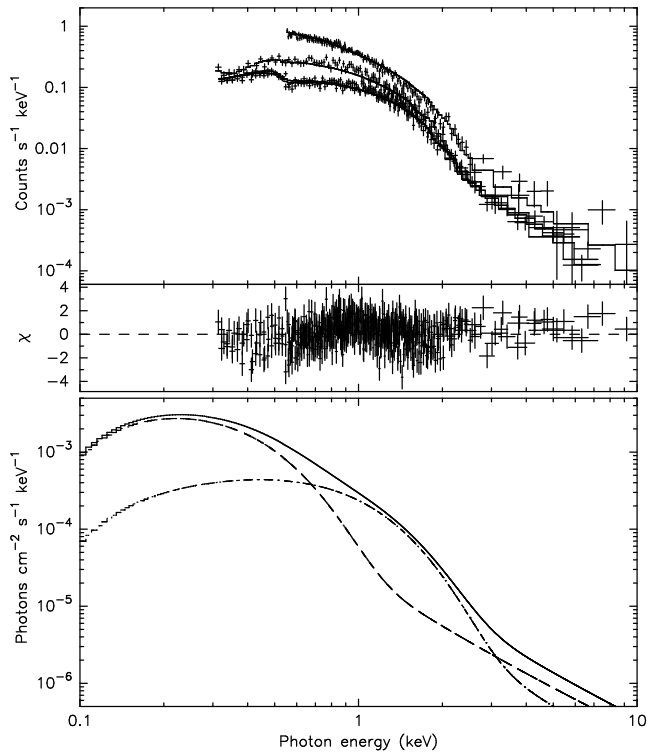} \\
\includegraphics[angle=270, width=0.47\textwidth]{f1b.ps}
\caption{(\textit{top}) \textit{Chandra} ACIS-S and
\textit{XMM-Newton} EPIC-MOS1/2 and EPIC-pn spectra of J0437 fitted
with a two temperature Comptonized blackbody model ({\tt compbb}). The
lower panel shows the best fit residuals. (\textit{bottom}) Model
spectra used in the fit. The solid line is the total spectrum, while
the dashed and dot-dashed lines show the individual emission
components (see text for best fit parameters).}
\end{center}
\end{figure}

\subsection{Thermal Emission}

In a purely thermal scenario the hard emission from J0437 may
originate from a very small hot spot on the polar cap with a radius of
order a few meters, radiating at temperatures in excess of $8\times
10^6$ K. However, no current MSP polar cap heating model predicts such
peculiarly high temperatures and small emission areas, making this
model rather implausible. The validity of this model can be tested by
deep observations at $\gtrsim$3 keV.

\subsection{Comptonized Thermal Emission}

The hard X-ray tail can be naturally produced by reprocessing of the
thermal emission via inverse Compton scattering (ICS) by an optically
thin ($\tau<1$) hot plasma. As a thermal photon emitted at the NS
surface propagates through the tenuous co-rotating plasma in the
pulsar magnetosphere, it may undergo ICS by energetic e$^{\pm}$.  In
this regime, repeated scatterings of the thermal seed photons result
in a PL distribution of photons \citep[see][]{Ryb79,Nish86}.  Due to
the low $\tau$ of the magnetospheric plasma, only a small fraction of
the thermal photons are scattered, resulting in a faint Comptonized
tail.

A major advantage of this model is that, unlike the case of a
non-thermal (magnetospheric) PL, it does not extend below $\sim$1 keV
(see Fig. 1), thus, aleviating any discrepancies with optical/UV data.
Therefore, based on the available observational evidence and the
relative simplicity of the Comptonization model, we deem this
interpretation the most plausible.


\begin{figure}[!t]
\begin{center}
\includegraphics[width=0.47\textwidth]{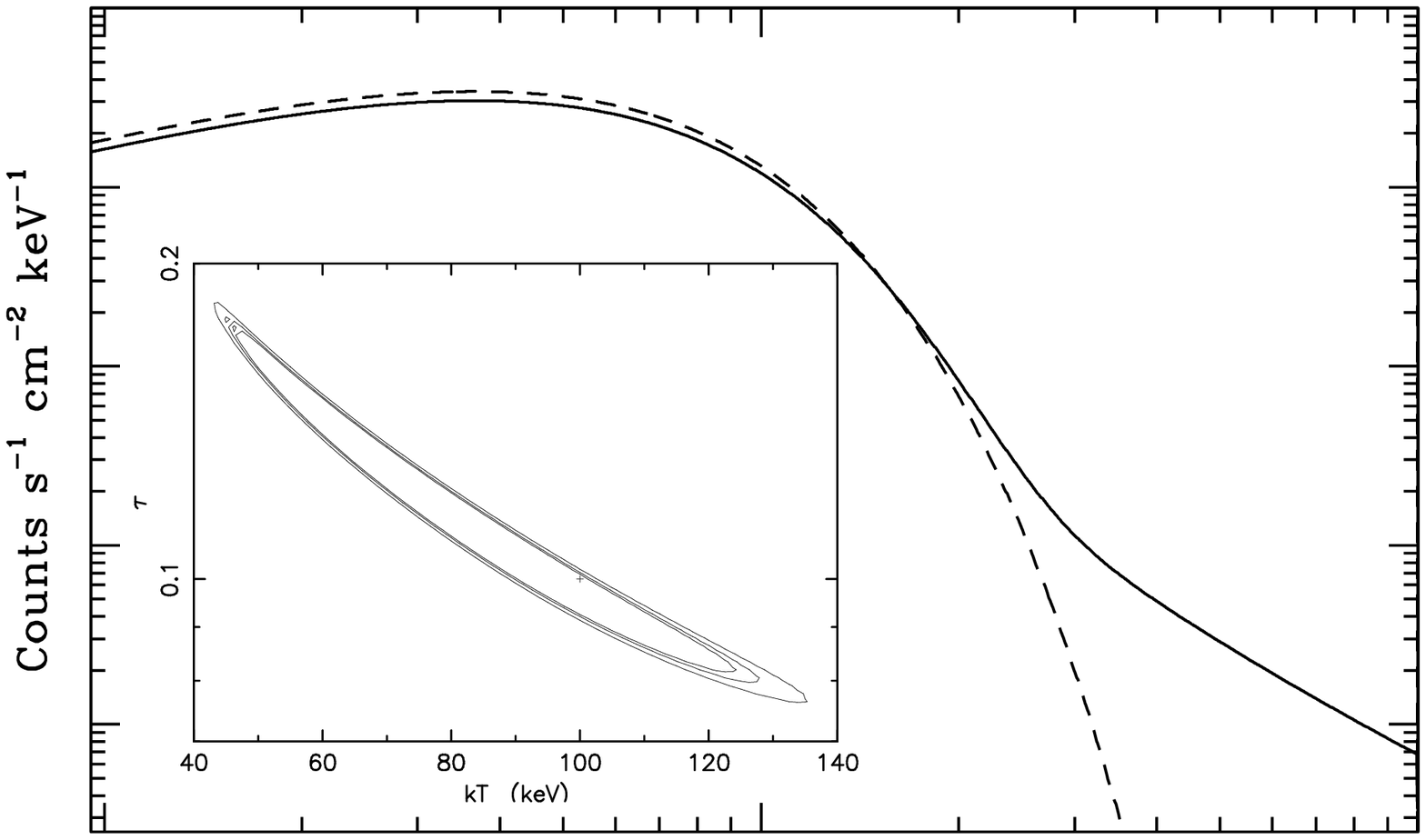} \\
\includegraphics[width=0.47\textwidth]{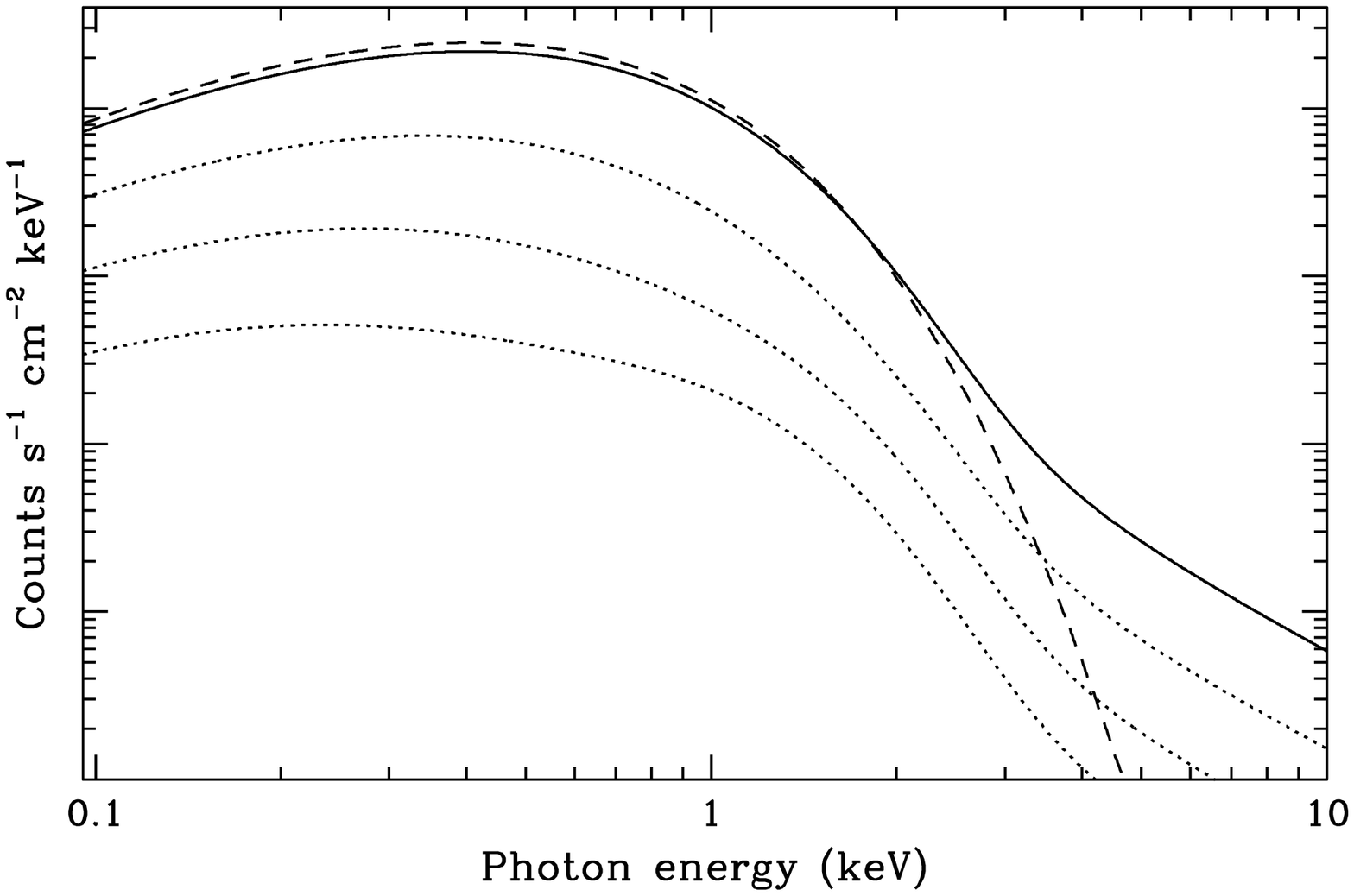}
\caption{(\textit{top}) Example Comptonized BB model spectrum for
$kT_{\rm eff}=0.23$ keV, $kT_e=100$ keV, and $\tau=0.1$. The dashed
line shows the initial thermal photon distribution. The inset shows
example 1-, 2-, and $3\sigma$ contours for $\tau$ versus $kT_e$ to
illustrate the correlation between the two
parameters. (\textit{bottom}) Comptonized NSA model \citep{Heinke06}
spectrum for $T_{\rm eff}=1.3\times10^6$ K, $kT_e=100$ keV, and
$\tau=0.1$. The photon energies have been corrected for the
gravitational redshift assuming $R=10$ km and $M=1.4$ M$_{\odot}$.
The dotted lines correspond to different viewing angles relative to
the surface normal ($\cos \theta = 0.5$, 0.25, and 0.1, from top to
bottom, respectively), while the solid line is for $\cos\theta=1$.  In
both plots the dashed line shows the initial thermal photon
distribution. The flux normalization is arbitrary.}
\end{center}
\end{figure}

\section{Results}

We have examined the effect of inverse Compton scattering (ICS) on the
thermal emission from MSPs for both a blackbody (BB) and neutron star
hydrogen atmosphere \citep[NSA, see][]{Rom87,Zavlin96} model.  For the
BB case we employed the {\tt compbb} model in XSPEC (Nishimura et
al. 1986).  This particular model assumes a plane-parallel,
semi-infinite scattering medium with the source of thermal photons at
the bottom. It is valid over the range $E\ll kT_e<mc^2$, where $E$ is
the initial photon energy, $T_e$ is the scattering e$^{\pm}$
temperature, and $mc^2$ is the electron rest energy.  Figure 1 shows
the X-ray spectrum of J0437 fitted with a two-temperature Comptonized
BB model using {\tt compbb}. Throughout the spectral fits we fixed the
hydrogen column density towards the MSP to $N_{\rm H}=2\times10^{19}$
cm$^{-2}$ and electron temperature to $T_e=150$ keV. The best fit
parameters we obtain are $T_1=(2.93\pm0.03)\times10^6$ K,
$R_1=44\pm11$ m, $\tau_1=0.06\pm0.02$, $T_2=(1.16\pm0.01)\times10^6$
K, $R_2=300\pm40$ m, and $\tau_2=0.09\pm0.01$. The corresponding
luminosity in the 0.3-10 keV is found to be
$L_X\approx3\times10^{30}$ ergs s$^{-1}$.  All uncertainties quoted
above are 1$\sigma$. The minimum value of $\chi_{\nu}^2=1.36$ (for 464
degrees of freedom) was found to be identical to that in the case of
the BB+BB+PL model. Given the uncertainties in the cross-calibration
of the different detectors, this value of $\chi_{\nu}^2$ indicates an
acceptable fit. As is generally the case for ICS, $T_e$ and $\tau$ are
strongly correlated (see inset in Fig. 2) so a good fit was obtained
for a wide range of $kT_e$ and $\tau$.

The {\tt compbb} model assumes a thermal distribution of scattering
e$^{\pm}$.  In reality, these particles probably follow a non-thermal
(PL) distribution over a certain range of energies. Notwithstanding,
the exact energy distribution of the e$^{\pm}$ has no bearing on the
validity of this model since in both cases, multiple scatterings
result in a PL distribution of photons.

To study the effect of Comptonization on the emission from a NSA we
used the unmagnetized hydrogen atmosphere model from Heinke et
al. (2006) (see also McClintock et al. 2004) and the Comptonization
algorithm of \citet{Nish86}.  As expected, we have found that the NSA
model yields a qualitatively similar spectrum to that of the BB case
(see Fig. 2).  However, crucial differences do exist, steming from the
inherently anisotropic and energy dependent emission pattern of a NSA
\citep[][]{Zavlin96}. In particular, at a given viewing angle, the
intensity of the NSA radiation is subject to limb darkening as well as
a projection effect. The scattered photons, on the other hand,
probably have a different angular distribution.  This suggests that
depending on the viewing geometry, the observed flux may show
variations in the relative contribution from the Comptonized
component.

It is important to emphasize that the exact geometry and location of
the scattering medium relative to the MSP polar caps is unknown and
cannot be determined from the time integrated spectra alone. In
principle, these properties can be inferred by studying the modulation
of the Comptonized radiation as a function of the rotational phase of
the MSP.  To investigate the temporal behavior of the PL tail, we used
the timing data from the \textit{XMM-Newton} EPIC-pn detector. Figure
3 shows lightcurves of J0437 folded at the MSP spin period ($P=5.76$
ms) for the 0.3-1 keV and 3-6 keV energy ranges. In the Comptonization
scenario, these two bands contain purely (unscattered) thermal and
just Comptonized photons, respectively.  Unfortunately, due to the
limited count statistics and the high instrument and sky backgrounds,
which dominate the emission for $>$3 keV, at present, we cannot
provide any meaningful constraints on the properties of the
Comptonizing medium. We are, however, able to place a limit on the
pulsed fraction of $\lesssim$50\% for $>$3 keV. This value further
constrains a non-thermal magnetospheric origin since such emission is
expected to be almost entirely pulsed.


\begin{figure}[!t]
\begin{center}
\includegraphics[width=0.47\textwidth]{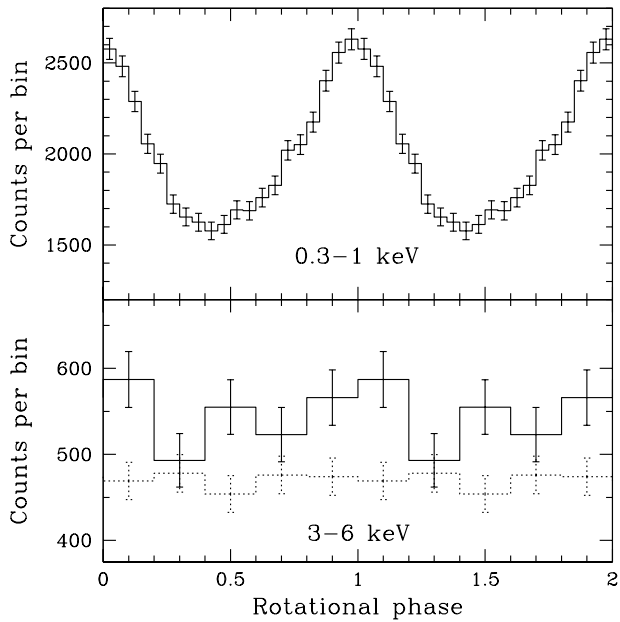}
\caption{\textit{XMM-Newton} EPIC-pn lightcurves of J0437 in the 0.3-1
keV (\textit{top}) and 3-6 keV (\textit{bottom}) bands. The dotted
line in the bottom panel shows the background level. Two cycles are
shown for clarity.}
\end{center}
\end{figure}

\section{DISCUSSION AND CONCLUSION}

We have demonstrated that the X-ray emission from the nearby MSP J0437
in the 0.1--10 keV band is well described by a two-temperature
Comptonized thermal spectrum.  This model provides an attractive and
relatively simple alternative to the non-thermal magnetospheric
emission model.  Since in the \citet{Hard02} model of high-energy
emission from MSPs, ICS is the primary mechanism for pair production
and polar cap heating, the presence of a hard Comptonized tail would
provide a strong confirmation that this mechanism does indeed operate
in MSP magnetospheres.

We note that the shape of the spectra in the non-thermal
magnetospheric and Comptonization cases are indistinguishable in the
0.1-10 keV band. The primary difference arises in the optical/UV range,
where only the former is expected. As mentioned previously, for J0437
the non-thermal model has been found to be only marginally consistent
with FUV observations.  The detection of a hard X-ray tail from
other near-by MSPs such as PSRs J0030+0451 and J2124--3358, would
provide a more definitive test of the two interpretations (the
existence of a hard X-ray component cannot be ascertained in present
X-ray observations of these MSPs due to the limited counts).  For
these solitary MSPs, the absence of a stellar companion has allowed
ultra-deep optical observations, which have found no emission down to
$V\sim28$ \citep{Kop03,Mig03}.

The Comptonization model proposed in this \textit{Letter} implies that
\textit{all} of the soft emission ($<$1 keV) from J0437-4715 is purely
thermal (unscattered) radiation.  This has profound ramifications in
the study of fundamental NS properties since thermal emission from the
surface of neutron stars reveals important information regarding the
basic properties of this class of compact objects.  Pavlov \& Zavlin
(1997) have shown that X-ray spectral and timing observations of MSPs
can be used to measure fundamental NS parameters such as the
mass-to-radius ratio ($M/R$) of the underlying NS. In addition, the
polar caps represent ``footprints'' of the pulsar magnetic field,
allowing constraints on the global field geometry.  Finally, since the
properties of the hard emission component are determined by the energy
distribution of scattering particles and their optical depth, study of
this radiation can also serve as a probe of the particles populating
the pulsar magnetosphere.  An attempt to constrain the properties of
J0437 outlined above is currently under way and will be presented in a
future paper.

\acknowledgements 
This work was supported in part by NASA grant
AR6-7010X. The research presented here has made use of the NASA
Astrophysics Data System (ADS).

%
%

\end{document}